\documentclass[showpacs,preprintnumbers,amsmath,amssymb]{revtex4}
\usepackage{graphicx}
\usepackage{dcolumn}
\usepackage{bm}
\begin{document}
\title{On the Poincar\'e and Gauge symmetry of a model where vector and
axial vector interaction get
 mixed up with different weight}
\author{Safia Yasmin}
\affiliation{Indas Mahavidyalaya, Bankura - 722205, West Bengal,
India}
\author{Anisur Rahaman} \email{1. anisur.rahman@saha.ac.in, 2.
manisurn@gmail.com} \affiliation{Hooghly Moghin College,
Chinsurah, Hooghly-712101, West Bengal, India}

\date{\today}

\begin{abstract}
A $(1+1)$ dimensional model where vector and axial vector
interaction get mixed up with different weight is considered with
a generalized masslike term for gauge field. Through Poincar\'e
algebra it has been made confirm that only a Lorentz covariant
masslike term leads to a physically sensible theory as long as the
number of constraints in the phase space is two. With that
admissible masslike term, phase space structure of this model with
proper identification of physical canonical pair has been
determined using Diracs' scheme of quantization of constrained
system. The bosonized version of the model remains gauge
non-invariant to start with. Therefore, with the inclusion of
appropriate Wess-Zumino term it is made gauge symmetric. An
alternative quantization has been carried out over this gauge
symmetric version to determine the phase space structure in this
situation. To establish that the Wess-Zumino fields allocates
themselves in the un-physical sector of the theory an attempts has
been made to get back the usual theory from the gauge symmetric
theory of the extended phase-space without hampering any physical
principle. It has been found that the role of gauge fixing is
crucial for this transmutation.
\end{abstract}

 \maketitle

\section{{\bf Introduction}}
In terms of  fundamental interaction, Quantum Electrodynamics
(QED) in (1+1) dimension  can be categories in two different
classes. The first way of description that came in the
 literature was originated from vector type of interaction between matter and gauge fields.
 The models which belong to this class are well known vector Schwinger
 model \cite{SCH} and  Thiring-Wess  model \cite{THIR}. The other way of description
  originated from chiral interaction between matter and
gauge fields. Chiral Schwinger model \cite{JR} along with its
different variants \cite{KH, PM, SG, SM1, SM2} and Chiral
 Thiring-Wess model \cite{ARANN1, ARANN2} are the example of this class. In the Chiral Schwinger model
  \cite{JR} and in its different variants \cite{KH, PM, SG, SM1, SM2},
  we find that Vector and axial vector interaction get mixed up with equal
  weight.
  Few years ago, the authors in \cite{BAS1} presented a model where unlike Chiral Schwinger
  model,
    vector and axial vector interactions did not mix up with equal weight. Few extensions over this model is also
    found in
     \cite{BAS2, BAS3}. The mixing
    of interaction with different weight may be regarded as a generalized version of QED (GVQED) which
     covers all the fundamentally different interaction and their mixing \cite{SCH, THIR, JR, KH, PM}.
      The beauty of this
     model is that it is capable of interpolating both the QED and chiral QED. Both the Schwinger
      model \cite{SCH} and the Chiral Schwinger model \cite{JR} can be achieved through the
      different choice its mixing
      weight factor of
      interaction. For unit weight factor it describes the Chiral Schwinger model \cite{JR} and for vanishing
       weight it describe the vector Schwinger model \cite{SCH}.

        Standard quantization scheme furnishes that
       these two models are fundamentally different so far theoretical spectrum and confinement aspect of fermion
       are concerned \cite{LO, ROT1, ROT2}. Needless to mention that Schwinger model \cite{SCH}, and its
       chiral generation, e.g.,
        Chiral Schwinger
        model \cite{JR, CRH}, and the GVQED as presented by in \cite{BAS1, BAS2, BAS3}, which covers the
        both into its own are of considerable interest because of their ability to describe different physical
         aspects which are found to exist even in (3+1) dimension. Schwinger model acquired
         popularity not only for its ability of describing mass generation via dynamical symmetry
         breaking \cite{SCH, LO}, but also it can describe the confinement property of fermion in lower
         dimension \cite{SCH, LO}
         which is a real (3+1) dimensional
          aspect of QCD. On the other hand, Chiral Schwinger model is capable of
describing mass generation as well like vector Schwinger model
\cite{JR}, however fermions are found to get liberated
           here which can be considered as lower dimensional de-confining state of fermion \cite{JR, KH, PM}. Since the
          GVQED  presented in \cite{BAS1, BAS2, BAS3}, interpolates both the Schwinger model
and Chiral
           generation of that, it is natural that all the surprises involved in the
           Schwinger model and Chiral Schwinger models lies significantly in this GVQED.
All these models along with the GVQED \cite{BAS1, BAS2} are so
rich
 in describing, different surprises like dynamical mass generation, confinement and de-confinement
  aspects of fermion, that till now investigation over these models are carried out
  \cite{ARANN1, ARANN2, AR1, AR2, AR3, AR4, AR5, AR6, AR7, AR8, AAP1, AAP2, ARS, CAS1, CAS2, MUS, MIC,
   ADAS, MIA, SADOO, GHAS, KUL} and
  these models still remains
   as a fertile field to carry out further investigations. Our objective in this work is to carry
   out few investigations over the GVQED coined in \cite{BAS1} concerning the
     Poincar\'e  and gauge symmetry. An attempt is also made here to single
     out the real physical
     canonical pairs embedded within the phase space of the system.
      It is true that a systematic quantization of this model is available in \cite{BAS1}, however the definite identification of real physical
canonical
        pairs lying within the phase space is found to be absent. In order
        to make it a compliment to the quantization part of the work \cite{BAS1}, again
         quantization of this model has been carried
out using Diracs' scheme of quantization of constrained system.
Besides, the gauge current of the model is anomalous
        which
        leads to a gauge non-invariant structure. So quantization of the gauge
        invariant version of this model
         would also be of considerable interest. In this
        context,
        gauge symmetric construction is
made here extending the phase space with the inclusion of
appropriate Wess-Zumino action \cite{WSJ} and in presence of that
Wess-Zumino term  an extension towards an alternative quantization
is also made here to determine the canonical
             pair of fields which describe the Fock-space.
         As a natural corollary,
          it is shown that the physical contents of the theory remains identical, even after the
           extension of phase space by the use of the technique available from the work of
            Falck and Kramer in \cite{FALCK}.
            It is shown here explicitly that an appropriate gauge fixing
           is capable of mapping the Wess-Zumino added action onto
           the initial gauge non-invariant
            effective action.  The
             plan of the paper is as follows.

             In Sec. II, through the Poincar\'e algebra
             investigation has been carried out towards the search of the appropriate structure
              of masslike term which will be able to lead to a physically
              sensible theory starting from a very
             generalized masslike term.  Sec. III,
             deals with the
             Diracs'
             scheme of quantization of constrained system where we have
             attempted to
             single out the
             the real physical canonical pairs in a transparent manner
             to make this part a complement
             to the work \cite{BAS1}. An
             alternative quantization of the gauge invariant
             version of this model is made in Sec. IV. In Sec. V,
             it is shown that an appropriate gauge fixing can map the gauge invariant theory of the
             extended phase space onto the usual gauge non-invariant
             structure of it.

\section{{\bf To find out the lorentz transformation of the fields and
evaluation of the requirement to be  physically sensible}}
  A model where we find the mixing of both vector and axial vector interaction with different
  weight is given by the following generating functional
\begin{equation}
Z(A)=\int{d\psi} {d\bar{\psi}} exp[i \int{d^2x}L_F]. \label{INT}
\end{equation}
with
\begin{equation}
L_{F}=\bar{\psi}\gamma^{\mu}[i\partial_{\mu}+e\sqrt{\pi}A_{\mu}(1-r\gamma_{5})]\psi.
\end{equation}
The integration over the fermionic degrees of freedom $\psi$ leads
to a determinant which is singular in nature \cite{BAS1, ROT1,
ROT2}. In order to remove the singularity we need to regularize
the theory. After proper regularization if we express
 the fermionic
determinant in terms of auxiliary scalar field $\phi$, we get
\begin{equation}
Z(A)=\int{d\phi} exp[i \int{d^2x}{\cal L}_B]. \label{INT1}
\end{equation}
with
\begin{equation}
{\cal L}_{B}=\frac{1}{2}
\partial_{\mu}\phi\partial^{\mu}\phi+eA^{\mu}(\tilde{\partial}_{\mu}+r\partial_{\mu})\phi+\frac{e^2}{2}
(\alpha A_0^2 + \beta A_0 A_1 + \gamma
A_1^2),\label{LB}\end{equation} where $\tilde{\partial}_{\mu}=
\epsilon_{\mu\nu}\partial^\nu$ and $\epsilon^{01} =+1$. A
generalized masslike term has been included here as counter term
in place of standard $\frac{1}{2}ae^2A_\mu A^\mu$ term since we
are intended to study whether any other alternative  masslike term
can serve as a physically sensible counter term for regularization
like the chiral Schwinger model \cite{PM, SG, SM1, SM2}. The
parameters $\alpha, \beta$ and $\gamma$, therefore, stand as the
regularization ambiguity parameter. Needless to mention that in
this situation ambiguity emerged out during the process of
regularization in order to remove the divergence of the fermionic
determinant. If we now take into account the kinetic term of the
back ground electromagnetic field the lagrange density  then turns
into
\begin{eqnarray}
{\cal L}_{B} &
=&\frac{1}{2}\partial_\mu\phi\partial^\mu\phi+eA^\mu(\epsilon_{\mu\nu}\partial^\nu
+rg_{\mu\nu}\partial^\nu)\phi\nonumber \\
 &+&
\frac {1}{2}e^2(\alpha A_{0}^{2}+2\beta A_{0}A_{1}+\gamma
A_{1}^{2})-\frac{1}{4}F_{\mu\nu}F^{\mu\nu}. \label{LAG}
\end{eqnarray}
Starting with this generalized masslike term we  now proceed to
investigate which type of term  leads to a physically sensible
theory.  The word physically sensible implies a structure that not
only maintains physical Lorentz invariance but also leads to an
exactly solvable nature at the same time. To this end, we would
like to study the Loentz transformation property of the fields and
the Poincar\'e algebra of the theory in an explicit manner
\cite{PM}. In this context, we need to calculate the momenta of
the fields describing the theory. From the standard definition the
momenta corresponding to the fields $\phi$ , $A_{0}$ and  $A_{1}$
are found out:
\begin{equation}
\pi_{\phi}=\dot{\phi}-eA_{1}+erA_{0}, \label{FFBR1}
\end{equation}
\begin{equation}
\pi_{0}=0, \label{FFBR2}
\end{equation}
\begin{equation}
\pi_{1}=\dot{A_{1}}-A_{0}^{\prime}. \label{FFBR3}
\end{equation}
Here $\pi_{\phi}$, $\pi_{0}$, and $\pi_{1}$ are the momenta
corresponding to the fields $\phi$, $A_0$ and $A_1$. For this
theory $\Omega_1=\pi_{0}\approx 0$, is the primary constraint. A
Legendre transformation leads to the the following canonical
Hamiltonian density.
\begin{eqnarray}
{\cal H}_{c}&=&\frac{1}{2}(\pi_{\phi}^{2}+\phi'^{2}
+\pi_{1}^{2})+\pi_{1}A_{0}^{\prime}+\frac{1}{2}e^{2}(A_{1}-rA_{0})^{2}\nonumber\\
&+&e\pi_{\phi}(A_{1}-rA_{0})-e(A_{0}\phi^{\prime}-rA_{1}\phi^{\prime})\nonumber\\
&-& \frac{e^{2}}{2}(\alpha A_{0}^{2}+2\beta A_{0}A_{1}+\gamma
A_{1}^{2}). \label{FFBR4}
\end{eqnarray}
Time evolution of primary constraint with respect to the
Hamiltonian gives the following secondary constraint:
\begin{equation}
\Omega_{2}=\pi'_1+e^2((\alpha-r^2)A_{0}+(r+\beta)A_1)+e(r\pi_\phi+e\phi')\approx
0 \label{FFBR5}
\end{equation}
The constraints are all weak conditions at this stage. To impose
it as a strong condition  into the system we need to have the
expression of $A_{0}$. Equation (\ref{FFBR5}), gives
\begin{equation}
A_{0}=-\frac{1}{e^{2}(\alpha-r^{2})}(\pi_{1}^{\prime}+e(r\pi_{\phi}+e\phi^{\prime})+e^{2}(r+\beta)A_{1}).
\label{FFBR6}
\end{equation}
Inserting the value of $A_{0}$ in equation (\ref{FFBR4})  we get
the following reduced Hamiltonian.
\begin{eqnarray}
H_{R}&=&\int[\frac{\pi_{1}^{2}}{2}+\frac{\pi_{1}^{\prime^{2}}}{2e^{2}(\alpha-r^{2})}
+\frac{1}{2}\frac{\alpha\pi_{\phi}^{2}}{(\alpha-r^{2})}
+\frac{e^{2}}{2}((1-\gamma)+\frac{{(\beta+r)}^{2}}{(\alpha-r^{2})})A_{1}^{2} \nonumber\\
&+&\frac{(1+\alpha-r^{2})}{(\alpha-r^{2})}\frac{\phi^{\prime^{2}}}{2}
+ \frac{(\beta+r+\alpha
r-r^{3})}{(\alpha-r^{2})}eA_{1}\phi^{\prime} + \frac{(\alpha+\beta
r)}{(\alpha-r^{2})}eA_{1}\pi_{\phi} \nonumber\\
 &+&
\frac{(\beta+r)}{(\alpha-r^{2})}\pi_{1}^{\prime}A_{1}
+\frac{r}{(\alpha-r^{2})}{\phi}^{\prime}\pi_{\phi}
+\frac{\phi^{\prime}\pi_{1}^{\prime}}{e(\alpha-r^{2})}
+\frac{r}{e(\alpha-r^{2})}\pi_{\phi}\pi_{1}^{\prime}].
\label{FFBR7}
\end{eqnarray}
For this reduced Hamiltonian the ordinary Poission's brackets
become inadequate \cite{DIR}. So it becomes essential to calculate
the Dirac brackets between the fields describing the Hamiltonian
to proceed further. The Dirac bracket \cite{DIR} between the two
variables $A$ and $B$ is defined by
 \begin{equation}
 [A(x), B(y)]^* = [A(x), B(y)] - \int[A(x) \omega_i(\eta)]
 C^{-1}_{ij}(\eta, z)[\omega_j(z), B(y)]d\eta dz, \label{DDIR}
 \end{equation}
 where $C^{-1}_{ij}(x,y)$ is given by
 \begin{equation}
 \int C^{-1}_{ij}(x,z) [\omega_i(z), \omega_j(y)]dz = 1.
 \label{INV} \end{equation}
 Here $\omega
 _i$'s represents the second class constraints that remains embedded within the phase space
 of the theory.
 The matrix $C^{-1}(x,y)$ for the theory under consideration is
\begin{equation}
C^{-1}(x, y) =
\frac{1}{e^2(\alpha-r^2)} \left(\begin{array}{cc} 0 & \delta(x-y)\\
                -\delta(x-y) & 0 \end{array}\right),
 \label{MAT}\end{equation}
Our task becomes little easier since it is found that the Dirac
brackets between the fields remains canonical.
\begin{equation}
[A_1(x), \pi_1(y)]^*= \delta(x-y), \label{DB1}
\end{equation}
\begin{equation}
[\phi(x), \pi_\phi(y)]^*= \delta(x-y),\label{DB2}
\end{equation}
\begin{equation}
[A_1(x), \phi(y)]^*= 0.\label{DB3}
\end{equation}
The reduced Hamiltonian can be expressed in the following form
after few steps of algebra:
\begin{eqnarray}
H_R &=&\int dx [ \frac{1}{2}(\pi_1^2+ \pi_\phi^2 +
\phi'^2)+\frac{1}{2}e^2(1-\gamma)A_1^2 \nonumber \\
&+&\frac{1}{2e^2(\alpha-r^2)}(\xi^2+ 2\partial_1 (\xi\pi_1)) +
eA_1(\pi_{\phi}+\phi')].
\end{eqnarray}
The total momentum and the boost generator in $(1+1)$ are defined
by
\begin{equation}
P=\int dx [\pi_1 A'_1, + \pi_0 A'_0 + \pi_\phi \phi'],\label{TM}
\end{equation}
and
\begin{equation}
M=t(\pi_{\phi}\phi^{\prime}+\pi_{1}A_{1}^{\prime}+ \pi_0 A'_0
)+\int dx[{xH_{R}}+\pi_{1}A_{0}+ \pi_0 A_1]. \label{BOOST}
\end{equation}
 In the reduced phase space that is in
the constrained subspace the equation (\ref{TM}) and (\ref{BOOST})
reads
\begin{equation}
P_R=\int dx [\pi_1 A'_1 + \pi_\phi \phi'], \label{CTM}
\end{equation}
\begin{equation}
M_R=tP_R+\int dx[x{\cal H}_R-\frac{1}{e^2(\alpha-r^2)}\pi_1 \xi]
\end{equation}
where
\begin{equation}
\xi=\pi'_1+e(r\pi_\phi+\phi')+e^2(r+\beta)A_{1},
\end{equation}
and the total Hamiltonian $H_R$ and the Hamiltonian density ${\cal
H}_R $ are related by $H_R = \int dx {\cal H}_R$. The momentum
operator $P_R$ transform the fields within the constrained
subspace. Similarly, the Hamiltonian operator $H_R$ generate the
time translation of the same. The time translation of the fields
are given by
\begin{eqnarray}
\dot{\phi}= [\phi(x), H_R(y)] =
\pi_{\phi}+eA_{1}+\frac{r}{e(\alpha-r^2)}\xi, \label{DOTP}
\end{eqnarray}
\begin{eqnarray}
\dot{A_{1}}=[A_1(x), H_R(y)]=
\pi_{1}-\frac{1}{e^{2}(\alpha-r^{2})}\xi'. \label{DOTA}
\end{eqnarray}
However, the most interesting one is the action of the
Lorentz-boost generator $M_R$ on the fields in the constrained
subspace. We now turn to observe that. Let us now see how the
fields get transformed under the Loentz-boost. Calculating the
Poisson brackets of the fields $\phi$ and $A_1$ with the
Lorentz-boost and expressing these in terms of $\dot\phi$ and
$\dot A_1$ using equation (\ref{DOTP}) and (\ref{DOTA}), we find
the expected transformation of the fields $\phi$ and $A_1$ under
the Lorentz-boost.
\begin{equation}
[\phi, M_{r}]=t\phi'+x\dot\phi. \label{LOP}
\end{equation}
\begin{equation}
[A_{1}, M_{R}]= tA'_1+x\dot A_1 +A_0. \label{LOA}
\end{equation}
With the use of the above transformation rules (\ref{LOP}) and
({\ref{LOA}), and the Dirac brackets (\ref{DB1}), (\ref{DB2})  and
(\ref{DB3}), it is straight forward to see that the following
Poincar\'e algebra
\begin{equation}
[H_R, M_R]^* = P_R,
\end{equation}
\begin{equation}
[P_R, M_R]^* = H_R,
\end{equation}
\begin{equation}
[P_R, H_R]^* = 0.
\end{equation}
is satisfied iff $\beta=0$ and $\alpha=-\gamma$. We should mention
here that it is valid only for the very structure of the
constraints  which are given in equation (\ref{FFBR2}) and
(\ref{FFBR5}). If we set $\alpha =r^2$, the constraint structure
will get altered and in that case total scenario will be
different. In fact, the number of constraint will be greater than
two in this situation like the Faddeevian \cite{FAD1, FAD2} class
of regularization of Chiral Schwinger model \cite{PM, SG, SM1,
SM2}. To study the aforesaid situation let us set  $\alpha=r^{2}$
and carry out the Poincar'/e algebra for this special case. The
constraint $\omega_{2}$ now takes the form
\begin{equation}
\bar\omega_{2}=\pi_{1}^{\prime}+e^{2}(r+\beta)A_{1}+er\pi_{\phi}+e\phi^{\prime}.
\label{exBRx5}
\end{equation}
The effective Hamiltonian of this theory in the present situation
can be written down as
\begin{equation}
H_{eff}=H+v\bar\omega_{2}+u\bar\omega_{1} \label{EXHEFF}
\end{equation}
The  consistency of  $\bar\omega_{2}$ with time requires
$\dot{\bar\omega}_{2} =0$, which fixes the velocity $v$. The
velocity $v$ is found out to
\begin{equation}
v=A_{0}+\frac{\gamma+r^{2}}{2\beta}A_{1}. \label{exBR89}
\end{equation}
With this velocity $v$ the $[\bar\omega_2(x), H(y)]$ gives birth
of a new constraint
\begin{equation}
\bar\omega_{3}=(r+\beta)\pi_{1}+2\beta{A_{0}^{\prime}}+(\gamma+r^{2})A_{1}^{\prime}
\end{equation}
So in the present situation, three constraints are embedded in the
phase space of the theory and the constraints are
\begin{equation}
\bar\omega_{1}=\pi_{0}, \label{EXCON1}
\end{equation}
\begin{equation}
\bar\omega_{2}=\pi_{1}^{\prime}+e^{2}(r+\beta)A_{1}+er\pi_{\phi}+e\phi^{\prime}\label{EXCON2}
\end{equation}
\begin{equation}
\bar\omega_{3}=(r+\beta)\pi_{1}+2\beta{A_{0}^{\prime}}+(\gamma+r^{2})A_{1}^{\prime}.\label{EXCON3}
\end{equation}
The matrix constructed out of the Poission's brackets within the
constraints is
\begin{eqnarray}
C_{ij}=\left(\begin{array}{ccc}
 0 & 0 & 2\beta\partial_1 \\
       0 & -2e^{2}\beta\partial_1 & (r^{2}+\gamma)\partial_1^{2}
       +e^{2}(r+\beta)^{2} \\
   2\beta\partial_1  & -(r^{2}+\gamma)\partial_1^{2}-e^{2}(r+\beta)^{2}& 2(r+\beta)(r^{2}+\gamma)\partial_1 \\

\end{array}\right)\delta(x-y)
\end{eqnarray}
The Hamiltonian in the reduced phase space in this situation reads
\begin{equation}
\bar H_{r}=\int dx[
\frac{(1+r^{2})}{r^{2}}\frac{\phi^{\prime^{2}}}{2}+\frac{\pi_{1}^{2}}{2}+\frac{1}{2e^{2}r^{2}}\pi_{1}^{\prime^{2}}
+\frac{e^{2}}{2}(\frac{\beta^{2}}{r^{2}}-\gamma)A_{1}^{2}+e(r+\frac{\beta}{r^{2}})A_{1}\phi^{\prime}
+\frac{1}{er^{2}}\phi^{\prime}\pi_{1}^{\prime}+\frac{\beta}{r^{2}}\pi_{1}^{\prime}A_{1}].
\label{EXHR}
\end{equation}
The dirac brackets of the fields with which the reduced
Hamiltonian is constituted with are computed as follows.
\begin{equation}
[A_{1}(x),A_{1}(y)]^*=\frac{1}{2e^{2}\beta}\delta^{\prime}(x-y),
\label{DAA}
\end{equation}
\begin{equation}
[A_{1}(x),\pi_{1}(y)]^*=\frac{(\beta+r)}{2\beta}\delta(x-y),
\label{DAPH}
\end{equation}
\begin{equation}
[\pi_{1}(x),\pi_{1}(y)]^*=\frac{e^{2}(r+\beta)^{2}}{4\beta}\epsilon(x-y),\label{DAPP}
\end{equation}
\begin{equation}
[\phi(x),\pi_{1}(y)]^*=-\frac{er}{4\beta}(r+\beta)\epsilon(x-y),
\label{DPHP}
\end{equation}
\begin{equation}
[\phi(x),\phi(y)]^*=\frac{r^{2}}{4\beta}\epsilon(x-y).\label{DPHPH}
\end{equation}
Let us now proceed to calculate the poincar\'e algebra for this
special situation. There are three elements in  this algebra like
the previous situation. One of the elements of course, is $\bar
H_{r}$, which is given in equation (\ref{EXHR}), and the rest of
the two are the total momentum and the boost generator. These two
respectively are
\begin{equation}
\bar P=\int dx [\pi_1 A'_1, + \pi_0 A'_0 + \pi_\phi
\phi'],\label{ETM}
\end{equation}
and
\begin{equation}
\bar M=t(\pi_{\phi}\phi^{\prime}+\pi_{1}A_{1}^{\prime}+ \pi_0 A'_0
)+\int dx[{xH_{R}}+\pi_{1}A_{0}+ \pi_0 A_1]. \label{EBOOST}
\end{equation}
In the constrained subspace space these two reduce to
\begin{equation}
\bar P_{r}=\pi_{\phi}\phi^{\prime}+\pi_{1}A_{1}^{\prime},
\label{exBR149}
\end{equation}
and
\begin{equation}
\bar M_{r}=t(\pi_{\phi}\phi^{\prime}+\pi_{1}A_{1}^{\prime})+
\int{xH_{R}}dx-\pi_{1}[\frac{1}{2\beta}(r+\beta)\partial^{-1}\pi_{1}+\frac{(\gamma+r^{2})}{2\beta}A_{1}].
\label{exBR249}
\end{equation}
respectively. The the action of the Lorentz-boost generator $\bar
M_R$ on the fields in the constrained subspace for this case is
\begin{equation}
[\phi, M_{r}]=t\phi'+x\dot\phi. \label{LOPH}
\end{equation}
\begin{equation}
[A_{1}, M_{R}]= tA'_1+x\dot A_1 +A_0. \label{LOA1}
\end{equation}
\begin{equation}
[A_{0}, M_{R}]= tA'_0+x\dot A_1 +A_1. \label{LOA0}
\end{equation}
With the use of the above transformation rules (\ref{LOPH}) and
({\ref{LOA1}), and the Dirac brackets (\ref{DAA}),(\ref{DAPH})
(\ref{DAPP}),(\ref{DPHP})  and (\ref{DPHPH}), a little algebra
shows that the following Poincar\'e algebra
\begin{equation}
[H_R, M_R]^* = P_R,
\end{equation}
\begin{equation}
[P_R, M_R]^* = H_R,
\end{equation}
\begin{equation}
[P_R, H_R]^* = 0.
\end{equation}
holds if the conditions $r^2=1$ and $2\beta+r(1+\gamma)=0$ are
satisfied simultaneously. This result agrees with result available
in \cite{PM, SG, SM1, SM2} for weight factor $r=-1$ with the
choice of parameters $\beta=-1$ and $\gamma=-3$. The result also
reminds the result obtained in \cite{ARANN2}. At this point we
would like to end up our the investigation through Poincar'/e
algebra on this model  and would like to proceed with the lorentz
covariant mass like term for the gauge field (which of course is
result obtained from the Poincar'/e algebra )and carry out
investigation to shed light on some of the important facts those
which would be of orth unravelling for this model.

\section {{\bf Singling out of the real physical canonical pair using Dirac quantization scheme}}
Putting $\beta=0$ and $\alpha=-\gamma$,(a condition for
maintenance of Lorentz invariance) and setting $\alpha=a$, we get
a lorentz covariant mass-like term for gauge field and the reduced
Hamiltonian with this setting reads
\begin{eqnarray}
H_{R}&=&\frac{\pi_{1}^{2}}{2}+\frac{\pi_{1}^{\prime^{2}}}{2e^{2}(a-r^{2})}
+\frac{a
\pi_{\phi}^{2}}{2(a-r^{2})}+\frac{e^{2}}{2}\frac{a(1+a-r^{2})}{(a-r^{2})}A_{1}^{2}\nonumber\\
&+& \frac{(1+a-r^{2})}{(a-r^{2})}\frac{\phi^{\prime^{2}}}{2}
+\frac{r{\phi}^{\prime}\pi_{\phi}}{(a-r^{2})}
+\frac{\phi^{\prime}\pi_{1}^{\prime}}{e(a-r^{2})}+\frac{r\pi_{\phi}\pi_{1}^{\prime}}{e(a-r^{2})}
\nonumber\\
 &+&er\frac{(1+a -r^{2})}{(a-r^{2})}A_{1}\phi^{\prime}
+ e\frac{a}{(a-r^{2})}A_{1}\pi_{\phi}
+\frac{r\pi_{1}^{\prime}A_{1}}{(a-r^{2})}. \label{FFBR8}
\end{eqnarray}
Using Dirac brackets (\ref{DB1}), (\ref{DB2}) and (\ref{DB3}), we
get the following first order differential equations of motion for
the fields describing the theory in the constrained subspace.
\begin{equation}
\dot{A_{1}}=\pi_{1}-\frac{1}{e^{2}(a-r^{2})}\pi_{1}^{\prime\prime}-\frac{r}{(a-r^{2})}A_{1}^{\prime}-
\frac{\phi^{\prime\prime}}{e(a-r^{2})}-\frac{r\pi_{\phi}^{\prime}}{e(a-r^{2})},
\label{FFBR9}
\end{equation}
\begin{eqnarray}
\dot{\pi_{1}}&=&-e^{2}a\frac{(1+a-r^{2})}{(a-r^{2})}A_{1}-er\frac{(1+a-r^{2})}{(a-r^{2})}\phi^{\prime}
-e\frac{a}{(a-r^{2})}\pi_{\phi}\nonumber \\
&-&\frac{r}{(a-r^{2})}\pi_{1}^{\prime},\label{FFBR10}
\end{eqnarray}
\begin{equation}
\dot{\pi_{\phi}}=\frac{(1+a-r^{2})}{(a-r^{2})}\phi^{\prime\prime}
+e\frac{r(1+a-r^{2})}{a-r^{2}}A_{1}^{\prime}
+\frac{r}{a-r^{2}}\pi_{\phi}^{\prime}+\frac{\pi_{1}^{\prime\prime}}{e(a-r^{2})},
\label{FFBR11}
\end{equation}
\begin{equation}
\dot{\phi}=\frac{a}{(a-r^{2})}\pi_{\phi}+\frac{ea}{(a-r^{2})}A_{1}+\frac{r}{e(a-r^{2})}\pi_{1}^{\prime}
+\frac{r}{(a-r^{2})}\phi^{\prime}.\label{FFBR12}
\end{equation}
A little algebra converts the above four equations (\ref{FFBR9}),
(\ref{FFBR10}), (\ref{FFBR11}) and (\ref{FFBR12}),  to the
following  second order differential equations.
\begin{equation}
[\Box+e^{2}\frac{a(1+a-r^{2})}{(a-r^{2})}]\pi_{1}=0,
\label{FFBR13}
\end{equation}
\begin{equation}
\Box[\phi+e\frac{1}{(1+a-r^{2})}\pi_{1}]=0, \label{FFBR14}
\end{equation}
\begin{equation}
[\Box+e^{2}\frac{a(1+a-r^{2})}{(a-r^{2})}](A_{1}+\frac{r}{ea}\phi')=0,
\label{FFBR15}
\end{equation}
\begin{equation}
\Box(\pi_{\phi}+ \frac{r}{ea}\pi_{1}^{\prime})=0. \label{FFBR16}
\end{equation}
Now a careful look reveals that within the above four equations
(\ref{FFBR13}), (\ref{FFBR14}), (\ref{FFBR15}),  and
(\ref{FFBR16}) the theoretical spectra  are hidden in a
significant manner. Note that, equation (\ref{FFBR13}), describes
a massive boson with square of the mass $m^2 =
e^{2}\frac{a(1+a-r^{2})}{(a-r^{2})}$ and equation (\ref{FFBR14})
describes a massless boson which is equivalent to a free fermion
in $(1+1)$ dimension. So unlike the Schwinger model, fermions gets
de-confined here. We have noticed that equation (\ref{FFBR15}) and
(\ref{FFBR16}), describe the Klein-Gordon type equations for a
massive and a massless excitation respectively. The fields
describing equations (\ref{FFBR15}) and (\ref{FFBR16}), can be
considered as the momenta corresponding to the fields satisfying
equation (\ref{FFBR13}) and (\ref{FFBR14}). Note that the fields
satisfying equation (\ref{FFBR13}) and (\ref{FFBR15}), satisfy
canonical poisson brackets between themselves. Similarly, the
fields satisfying equation (\ref{FFBR14}) and (\ref{FFBR16}),
satisfy the same canonical condition. So our description gives a
transparent picture not only for the theoretical spectrum but also
for the physical canonical pairs of the phase space. So this
section, will certainly complement the quantization part of the
work reported in \cite{BAS1}. Let us we end up the discussion
related to the theoretical spectra and identification of the real
canonical pair of the gauge non-invariant version of the GVQED and
proceed to deal with the gauge invariant version of the GVQED in
the next two sections.

\section{{\bf An alternative Quantization of the Gauge invariant version of
the theory in the extended phase space}}
The standard way of expressing a theory into its gauge invariant
version is to extend the phase space with the inclusion of
Wess-Zumino field \cite{WSJ}. So by adding the appropriate
Wess-Zumino action to the action of the usual bosonized gauge
non-invariant action we get a gauge invariant theory of the same
and the lagrangian of which is given by
\begin{eqnarray}
L&=& \int dx[\frac{1}{2}\partial_{\mu}\phi\partial^{\mu}\phi
+e\epsilon^{\mu\nu}A_{\mu}\partial_{\nu}\phi+erg^{\mu\nu}A_{\mu}\partial_{\nu}{\phi}
-\frac{1}{4}F_{\mu\nu}F^{\mu\nu}\nonumber\\
&+&\frac{ae^{2}}{2} A_{\mu}A^{\mu}
+\frac{1}{2}(a-r^{2})\partial_{\mu}\theta \partial^{\mu}\theta
-er\epsilon^{\mu\nu}A_{\mu}\partial_{\nu}\theta\nonumber\\
&+&e(a-r^{2})g^{\mu\nu}A_{\mu}\partial_{\nu}\theta+B\partial^{\mu}A_{\mu}
+ \frac{\tilde\alpha}{2}B^2 ]. \label{AQR}
\end{eqnarray}
The last two terms of the lagrangian (\ref{AQR}), imply the
lorentz type gauge fixing term at the action level. It is needed
for quantization in the alternative manner \cite{MIA1, MIA2, KHT,
ABDALLA}. The Euler-Lagrange equations of motion of the fields (of
both the usual and extended phase space) are
\begin{equation}
\Box{\phi}=-e\tilde{\partial_{\mu}}A^{\mu}-er\partial_{\mu}A^{\mu},
\label{AQR1}
\end{equation}
\begin{equation}
\Box\theta
=\frac{er}{(a-r^{2})}\tilde{\partial_{\mu}}A^{\mu}-e\partial_{\mu}A^{\mu},
\label{AQR3}
\end{equation}
\begin{equation}
\partial_{\mu}A^{\mu}+\tilde\alpha B=0, \label{AQR4}
\end{equation}
\begin{equation}
\partial_{\mu}F^{\mu\nu}-\partial^{\nu}B+J^{\nu}=0, \label{AQR5}
\end{equation}
where $J^{\mu}$ is the electromagnetic current which is defined by
\begin{equation}
J^{\mu}=e\epsilon^{\mu\nu}\partial_{\nu}\phi+erg^{\mu\nu}\partial_{\nu}\phi
-er\epsilon^{\mu\nu}\partial_{\nu}\theta+e(a-r^{2})g^{\mu\nu}\partial_{\nu}+e^2a
A^{\mu}.   \label{AQR6}
\end{equation}
 The  equation (\ref{AQR1}),(\ref{AQR3}),(\ref{AQR4}) and
 (\ref{AQR5}),
 agrees with
 the following exact solution of the  fields $\phi$, $\theta$
  and $A_{\mu}$.
 \begin{equation}
\phi=\frac{(a-r^{2})}{ea(1+a-r^{2})}F+\frac{h}{a}+r\eta,
\label{AQR7}
\end{equation}
\begin{equation}
\theta=-\frac{r}{{ea}{(1+a-r^{2})}}F-\frac{r}{{a}(a-r^{2})}h+\eta,
\label{AQR8}
\end{equation}
\begin{equation}
A_{\mu}=\frac{1}{e^{2}{a}}[{\frac{(a-r^{2})}{(1+a-r^{2})}}\tilde{\partial_{\mu}}F+
\partial_{\mu}B+e\tilde{\partial_{\mu}}h-ea{\partial_{\mu}}\eta],
\label{AQR9}
\end{equation}
where the fields $h,B,\eta,F$ are Fock-space fields. Equations
(\ref{AQR1}),(\ref{AQR3}),(\ref{AQR4}),(\ref{AQR5}),(\ref{AQR6}),(\ref{AQR7}),(\ref{AQR8})
and (\ref{AQR9}) after a few steps of algebra lead to the
following differential equations.
\begin{equation}
(\Box+m^{2})\Box{F}=0,  \label{AQR12}
\end{equation}
\begin{equation}
\Box{h}=0, \label{AQR13}
\end{equation}
\begin{equation}
\Box{B}=0,   \label{AQR14}
\end{equation}
\begin{equation}
\Box{\eta}=\tilde\alpha eB, \label{AQR16}
\end{equation}
where square of the mass is $m^2$ is given by
\begin{equation}
m^{2}=\frac{e^{2}a(1+a-r^{2})}{(a-r^{2})}. \label{AQR15}
\end{equation}
which is identical to the physical mass as we have obtained in
Sec. III, during the quantization of the system in its gage
non-invariant version. The Fock-space fields has the following
relation with canonical variables of the physical system.
\begin{equation}
\eta=\frac{a-r^2}{a}\phi+\frac{r}{a }\phi, \label{AQR17}
\end{equation}
\begin{equation}
h=(a-r^{2})(\phi-r\theta) -\frac{1}{e(1+a-r^{2}}\pi_{1}.
\label{AQR18}
\end{equation}
\begin{equation}
B=\pi_{0}.  \label{AQR19}
\end{equation}
\begin{equation}
F=\Box^{-1}\pi_{1}  \label{AQR20}
\end{equation}
Since $\pi_1 = -\epsilon^{\mu\nu}\partial_\nu A_\mu
=-\tilde\partial^\mu A_\mu$. It is straightforward to see that the
equal time commutator of the Fock-space fields are
\begin{equation}
 [\eta(x),\dot{\eta}(y)]=i\frac{1}{a}\delta(x-y),  \label{AQR21}
 \end{equation}
 \begin{equation}
  [F(x),\dot{F}(y)]=im^{2} \delta(x-y), \label{AQR22}
 \end{equation}
 \begin{equation}
 [h(x),\dot{h}(y)]=i\delta(x-y),   \label{AQR23}
 \end{equation}
 \begin{equation}
 [B(x),\dot{\eta}(y)]=ie\delta(x-y).  \label{AQR24}
 \end{equation}
This completes the quantization of the gauge invariant version of
the  theory in the extended phase space. When the phase space of a
theory is extended in order to restore the gauge symmetry it is
expected that the fields needed for the extension will allocate
themselves in the un-physical sector the theory. So it would be
interesting if get back the usual gauge non-invariant version from
the gauge symmetric one of the extended phase space peeping the
physical principles intact. We will now turn towards that.

\section{Appropriate gauge fixing to land onto the gauge non-invariant model from
its gauge invariant version}
In \cite{FALCK}, we have found a technique how to get back  the
usual gauge non-invariant theory from a gauge symmetric theory of
the extended phase space. We would like to make an extension  for
this GVQED following the guideline available in \cite{FALCK}
towards getting back the original gauge non-invariant theory. Let
us see how this technique respond to GVQED. Lagrangian of GVQED
when added with the Wess-Zumino term in order to restore the local
gauge symmetry turns into
\begin{eqnarray}
L_e&=&\int dx[
\frac{1}{2}(\dot{\phi^{2}}-\phi^{\prime^{2}})+\frac{1}{2}ae^2(A_0^2-A_1^2)
+\frac{1}{2}(\dot{A_{1}}-A_{0}^{\prime})^{2}\nonumber\\
&+&e(A_{0}\phi^{\prime}-A_{1}\dot{\phi})+er(A_{0}\dot{\phi}-A_{1}\phi^{\prime})
+\frac{1}{2}(a-r^2)(\dot{\theta^{2}}-\theta^{\prime^{2}})\nonumber\\
&-&er(A_{0}\theta^{\prime}-A_{1}\dot{\theta})
+e(a-r^{2})(A_{0}\dot{\theta}-A_{1}\theta^{\prime})]. \label{SBRS}
\end{eqnarray}
 Let us now proceed to calculate the momenta corresponding to the field  $ A_{0},A_{1},\phi$,and $\theta$. From the
 standard definition, the momenta corresponding to the
fields $ A_{0},A_{1},\phi$,and $\theta$ are found out:
\begin{equation}
\pi_{0}=0, \label{SBRS1}
\end{equation}
\begin{equation}
\pi_{1}=\dot{A_{1}}-A_{0}^{\prime}, \label{SBRS2}
\end{equation}
\begin{equation}
\pi_{\phi}=\dot{\phi}-eA_{1}+erA_{0},  \label{SBRS3}
\end{equation}
\begin{equation}
\pi_{\theta}= (a-r^{2})\dot{\theta}+e(a-r^{2})A_{0}+erA_{1}.
\label{SBRS4}
\end{equation}
Using (\ref{SBRS1}),(\ref{SBRS2}),(\ref{SBRS3}) and (\ref{SBRS4}),
canonical Hamiltonian in this situation is found out.
\begin{eqnarray}
H_{ce}&=& \int dx[\frac{1}{2}(\pi_\phi^{2}+\pi_1^2
+\phi'^2)+\pi_1A'_0+\frac{e^{2}a(1+a-r^{2})}{2(a-r^{2})}A_1^2
+e\phi'(rA_1-A_0)\nonumber\\
&+& e\pi_\phi(A_1-rA_0) +\frac{1}{2}(a-r^2)\theta'^2
+e\theta'((a-r^2)A_{1}-A_{0}) +\frac{1}{2(a-r^2)}{\pi_{\theta}^{2}}\nonumber\\
&+&\frac{e}{(a-r^{2})}(rA_{1}+(a-r^{2})A_{0})\pi_{\theta}].
\label{SBRS5}
\end{eqnarray}
Equation (\ref{SBRS1}) is independent of velocity. So it is the
primary constraint of the theory as usual. The time evolution of
the primary constraint (\ref{SBRS1}), with respect to the
Hamiltonian is
\begin{equation}
[\pi_{0},H_{ce}]=\pi_{1}^{\prime}+e(er\pi_\phi+
\phi')-e(\pi_\theta - r\theta')\approx 0, \label{SBRS6}
\end{equation}
which gives the secondary constraint of the theory. It is found
that the Poission bracket of the secondary constraint
$\tilde\omega_{2}$ with the Hamiltonian vanishes. So there lies
only two constraints in the phase space of the theory. Following
are those two.
\begin{equation}
\tilde\omega_{1}=\pi_{0}\approx 0, \label{SBRS7}
\end{equation}
\begin{equation}
\tilde\omega_{2}= \pi_{1}^{\prime}+e(r\pi_\phi+
\phi')-e(\pi_\theta-r\theta')\approx 0 . \label{SBRS8}
\end{equation}
As it has been found in \cite{FALCK}, we too have introduced two
gauge conditions to get back the gauge non-invariant theory. These
two gauge fixing conditions are
\begin{equation}
\tilde\omega_{3}=\theta'\approx 0, \label{SBRS9}
\end{equation}
\begin{equation}
\tilde\omega_{4}=\pi_{\theta}+e((a-r^{2})A_{0}+rA_{1})\approx 0.
\label{SBRS10}
\end{equation}
Inserting the conditions  (\ref{SBRS9}) and (\ref{SBRS10},) as
strong condition into $\tilde\omega_{2}$ and $H_{ce}$, we find
that $\tilde\omega_{2}$ and $H_{ce}$,  reduce to the following
\begin{equation}
\tilde
\omega_{2R}=\pi_{1}^{\prime}+e^{2}(a-r^{2})A_{0}+e^{2}rA_{1}+e(r\pi_{\phi}+\phi')\approx
0 , \label{SBRS11}
\end{equation}
\begin{eqnarray}
\tilde H_R&=&\frac{1}{2}(\pi_1^2+\phi'^2)+\pi_1A_0'
+\frac{1}{2}ae^2 (A_1^2-A_0)^2\nonumber\\
&+&e\phi'(rA_1-A_0)+\frac{1}{2}[\pi_\phi+e(A_1-rA_0)]^2.\label{SBRS12}
\end{eqnarray}
Note that equation (\ref{SBRS11}) and  (\ref{SBRS11}), is
identical to the equations (\ref{FFBR5}) and (\ref{FFBR4}),
respectively, when $\alpha = -\gamma = a$ and $\beta=0$. For this
$\tilde H_R$, the ordinary poisson brackets become adequate
\cite{DIR}. So we need to evaluate the dirac brackets among the
fields. It necessities the computation of the matrix formed out of
the poisson brackets between the constraints along with the gauge
fixing conditions themselves. The constraints along with the gauge
fixing conditions gives the following matrix when Poisson brackets
among themselves are evaluated.
\begin{eqnarray}
C_{ij}=\left(\begin{array}{cccc}
0 & 0 & 0 & -e^{2}(a-r^{2}) \\
       0 & 0 & -e\partial_1 & 0 \\
       0 & -e\partial_1 & 0 & e\partial_1 \\
       e^{2}(a-r^{2}) & 0 & e\partial_1 & 0
\end{array}\right)\delta(x-y).
\end{eqnarray}
The determinant of $C_{ij}$ is non vanishing. So  it is invertible
and  its inverse is
\begin{eqnarray}
C_{ij}^{-1}= \frac{1}{e^2(a-r^2)}\left(\begin{array}{cccc}
    0 & \delta(x-y) & 0 & \delta(x-y)\\
     \delta(x-y) & 0 & -\frac{c}{2e}\epsilon{(x-y)} & 0\\
      0 & -\frac{c}{2e}\epsilon{(x-y)} & 0 & 0 \\
      -\delta(x-y) & 0 & 0 & 0,
     \end{array}\right)
      \end{eqnarray}
where the constant $c =e^2(a-r^2) $.
 Therefore, from the definition of Dirac brackets (\ref{DDIR}) the Dirac brackets between
the field variables can now be  computed in a straightforward
manner.
      \begin{equation}
[A_{0}(x),A _{1}(y)]^*=\frac{1}{e^{2}(a-r^{2})}\delta'(x-y),
\label{SBRS13}
\end{equation}
\begin{equation}
      [A_{0}(x),\phi(y)]^*=\frac{r}{e(a-r^{2})}\delta(x-y), \label{SBRS14}
      \end{equation}
      \begin{equation}
        [A_{0}(x),\pi_{\theta}(y)]^*=\frac{1}{2e}\epsilon{(x-y)}, \label{SBRS15}
            \end{equation}
            \begin{equation}
                [A_{0}(x),\pi_{\phi}(y)]^*=\frac{1}{e(a-r^{2})}\delta'(x-y), \label{SBRS16}
                 \end{equation}
                  \begin{equation}
                  [A_{1}(x),\pi_{\theta}(y)]=-\frac{1}{2e}\epsilon{(x-y)}, \label{SBRS17}
                  \end{equation}
                   \begin{equation}
                  [A_{0}(x),\pi_{1}(y)]^*=\frac{r}{(a-r^{2})}\delta(x-y), \label{SBRS18}
                  \end{equation}
                  \begin{equation}
                  [\pi_{\phi}(x),\pi_{\theta}(y)]^*=r\delta'(x-y), \label{SBRS19}
                  \end{equation}
                  \begin{equation}
                  [A_{1}(x),\pi_{1}(y)]^*=\delta(x-y), \label{SBRS20}
                  \end{equation}
                  \begin{equation}
                  [\phi(x),\pi_{\phi}(y)]^*=\delta(x-y). \label{SBRS21}
                  \end{equation}
 Note that the role of gauge fixing is very crucial role to gate back the usual theory since the other
  choice of valid  gauge fixing certainly exists, but that will
  lead to a different effective theory which may not help to get back to the usual
  theory in a straightforward manner.

\section{{\bf Conclusion}}
We have considered the GVQED coined in \cite{BAS1}, with a
generalized masslike term for gauge fields. It is added as a
counter term  to remove the divergence of the fermionic
determinant. In this context, we should mention that all possible
masslike term are not admissible as it gets restricted in order to
be  physically sensible, however massslike term may take some
generic shape. It may even take a structure which looks Lorentz
non-covariant however it does not stand as a hindrance in the way
of the theory to be exactly physical Lorentz invariant\cite{PM,
SG, SM1, SM2}. In this respect, an investigation through the
Poincar\'e algebra
 has been carried out using a generalized masslike
term for the gauge field. The algebra has imposed some condition
on the parameters used in the generalized masslike term and on the
weight factor of mixing. In fact, we have found two possibilities.
In the first case it does not put any restriction on the weight
factor of mixing, however it suggests a restriction that admits
the Lorentz covariant structure of the masslike term. No other
masslike term is admissible for this theory as long as its phase
space contains two constraints. In the second case, i.e., when
$\alpha =r^2$, it imposes restrictions on both the weight factor
and the parameters within the masslike term simultaneously. We
have found that the number of constraint in this situation is more
than two and the masslike term is of Lorentz non-covariant in
nature. It is worth mentioning here that the mixing weight $r\ne
1$ fails provide any physically sensible theory having  Lorentz
non-covariant masslike term.

With the admissible masslike term in the first possibility as
obtained from the Poincar\'e algebra, we quantize the theory using
the Dirac's scheme of quantization of constrained system. The
result though was known from the work available in \cite{BAS1}
that the theoretical spectrum contains a massive and massless
boson, nevertheless a more transparent calculation has been
presented here with the identification of real canonical pairs of
the phase space. Massive boson  as usual can be considered as
photons acquire mass via a dynamical symmetry breaking. On the
other hand, the massless boson of the theoretical spectrum may be
considered as free fermion. So fermion gets liberated here which
can be thought of as de-confinement in lower dimension. So the
model may be useful to study the lower dimensional QGP phenomena.
The quantization of the theory with masslike term as obtained in
the second possibility can get a ready idea form the work of one
od us \cite{ARANN2}, with few redefinitions of the parameter used
there.

The bosonized version of the model with the admissible masslike
term for the gauge fields obtained from both the possibilities are
found to be  gauge nonsymmetric. The first possibility is
considered here and it has been made gauge invariant with the
inclusion of Wess-Zumino field \cite{WSJ}, and an alternative
quantization of this gauge invariant version of the model is
carried out using lorentz gauge. The spectrum here too suggests
the appearance of the same missive and a mass less boson. Equation
(\ref{AQR12}) and (\ref{AQR13}), represents the massive and
massless field respectively. The extra equations (\ref{AQR14}) and
(\ref{AQR16}), appears because of the presence of the auxiliary
field $B$ in the Lorentz type gauge fixing term at the action
level. This type of investigation can be carried out for the
second possibilities also which we would like to carry out in our
future works.

When  phase space has been extended with the inclusion of
Wess-Zumino field to restore the gauge symmetry of the theory
generated from the fist possibility, it has been found that the
fields inserted for this extension allocate themselves in the
un-physical sector of the theory without disturbing the physical
sector at all. Using the method developed by Falck and Kramer  in
\cite{FALCK}, it has been found that an appropriate gauge fixing
maps the gauge invariant theory of the extended phase space onto
the usual gauge non-invariant theory. The role of gauge fixing is
found to be crucial here.

{\bf Acknowledgement}: One of us AR would like to thanks the
Director of Saha Institute of Nuclear Physics for his kind
permission to use the computer and library facilities of the
Institute. AR would also like to thank Prof. P. Mitra for a useful
discussion.

\end{document}